\documentclass[fleqn,usenatbib]{mnras}

\usepackage{newtxtext,newtxmath}

\usepackage[T1]{fontenc}
\usepackage{ae,aecompl}

\usepackage{units}
\usepackage{graphics,longtable}
\usepackage{graphicx,epsfig}
\usepackage{xcolor,xspace}

\newcommand{\kms}{km\,s$^{-1}$}

\renewcommand{\ion}[2]{{\textrm{#1}}{\textrm{\sc #2}}}
\newcommand{\mgd}{{$\mu$\textit{G3D}}\xspace}
\newcommand{\ourobj}{SDSS\,J0838\xspace}

\newcommand{\lya}{\relax \ifmmode {\mbox H}\alpha\else Ly$\alpha$\fi}
\newcommand{\ha}{\relax \ifmmode {\mbox H}\alpha\else H$\alpha$\fi}
\newcommand{\hb}{\relax \ifmmode {\mbox H}\beta\else H$\beta$\fi}
\newcommand{\hii}{\relax \ifmmode {\mbox H\,{\scshape ii}}\else H\,{\scshape ii}\fi}
\newcommand{\sii}{\relax \ifmmode {\mbox S\,{\scshape ii}}\else S\,{\scshape ii}\fi}
\newcommand{\nii}{\relax \ifmmode {\mbox N\,{\scshape ii}}\else N\,{\scshape ii}\fi}
\newcommand{\oi}{\relax \ifmmode {\mbox O\,{\scshape i}}\else O\,{\scshape i}\fi}
\newcommand{\oii}{\relax \ifmmode {\mbox O\,{\scshape ii}}\else O\,{\scshape ii}\fi}
\newcommand{\oiii}{\relax \ifmmode {\mbox O\,{\scshape iii}}\else O\,{\scshape iii}\fi}
\newcommand{\heii}{\relax \ifmmode {\mbox He\,{\scshape ii}}\else He\,{\scshape ii}\fi}

\title[Integral Field Spectroscopy of Green Peas (I)]{Integral Field Spectroscopy of Green Peas (I): Disentangling disk-like, turbulence and strong outflow kinematics in SDSSJ083843.63+385350.5}

\author[Bosch et al.]{G.\ Bosch$^{1}$, G.\ F.\ H\"agele$^{1}$, R.\ Amor\'in$^{2,3}$, V.\ Firpo$^{4,3}$, M.\ V.\ Cardaci$^{1}$, \and J.\ M.\ V\'ilchez$^5$, E.\ P\'erez-Montero$^5$, 
P.\ Papaderos$^6$, O.\ L.\ Dors$^7$, A.\ C.\ Krabbe$^7$, \and F.\ Campuzano-Castro$^{1}$\ \\
$^{1}$Instituto de Astrof\'isica de La Plata (UNLP - CONICET), La Plata, Argentina\\
$^{2}$Instituto de Investigaci\'on Multidisciplinar en Ciencia y Tecnolog\'ia, Universidad de La Serena, Ra\'ul Bitr\'an 1305, La Serena, Chile\\
$^{3}$Departamento de F\'isica y Astronom\'ia, Universidad de La Serena, Av. Juan Cisternas 1200 Norte, La Serena, Chile\\
$^{4}$Gemini Observatory, Southern Operations Center, La Serena, Chile\\
$^{5}$Instituto de Astrof\'isica de Andaluc\'ia - CSIC, Glorieta de la Astronom\'ia s.n., E-18008 Granada, Spain\\
$^{6}$Centro de Astrof\'isica and Faculdade de Ci\^encias, Universidade do Porto, Rua das Estrelas, 4150-762, Porto, Portugal\\
$^{7}$Universidade do Vale do Para\'iba, Av. Shishima Hifumi, 2911, Cep
12244-000, S\~ao Jos\'e dos Campos, SP, Brazil
}

\date{Accepted XXX. Received YYY; in original form ZZZ}

\pubyear{2018}

\begin{document}
\label{firstpage}
\pagerange{\pageref{firstpage}--\pageref{lastpage}}
\maketitle

\begin{abstract}
Integral Field Spectroscopy (IFS) is well known for providing detailed insight of extended sources thanks to the possibility of handling space resolved spectroscopic information. Simple and straightforward analysis such as single line fitting yield interesting results, although it might miss a more complete picture in many cases. Violent star forming regions, such as starburst galaxies, display very complex emission line profiles due to multiple kinematic components superposed in the line of sight. We perform a spatially resolved kinematical study of a single Green Pea (GP) galaxy, SDSSJ083843.63+385350.5, using a new method for analyzing Integral Field Unit (IFU) observations of emission line spectra. The method considers the presence of multiple components in the emission-line profiles and makes use of a statistical indicator to determine the meaningful number of components to fit the observed profiles.
We are able to identify three distinct kinematic features throughout the field and discuss their link with a rotating component, a strong outflow and a turbulent mixing layer. We also derive an updated star formation rate for \ourobj and discuss the link between the observed signatures of a large scale outflow and of the Lyman continuum (LyC) leakage detected in GP galaxies. 
\end{abstract}




\section{Introduction}
\label{intro}


Low-mass galaxies experiencing a major phase of star formation, such as compact H{\sc ii} galaxies \citep{Terlevich-Melnick1981} and Green Peas \citep[GPs,][]{Cardamone+09}, are exceedingly rare in the local universe. For their properties, GP galaxies belong to a wider family called Extreme Emission Line Galaxies \citep{Amorin+15}, that includes a large range of redshift. Since GPs are relatively close objects, they are unique laboratories to study massive star formation and feedback under extreme physical conditions which closely resemble those of ultraviolet (UV) selected high redshift galaxies \citep[e.g.,][]{2016ApJ...830...52E,2016ApJ...825...41V, Amorin+17}. In particular,
located at redshifts $z \sim 0.1-0.3$, the GPs are extreme emission line galaxies identified by their unresolved appearance and greenish colors in the Sloan Digital Sky Survey \cite[SDSS,][]{York+00} color composite images due to their strong nebular emission line spectrum --mostly dominated by [\oiii] emission with equivalent widths (EW) exceeding several hundredths of \AA. The observed GP properties indicate a brief and young burst of star formation with very large specific star formation rates (sSFR), as implied from their high ongoing star formation rates and low stellar masses \cite[$\log(M_*/M_{\odot})\!<\!10$;][]{Cardamone+09,Amorin+12b,Izotov2011}. 
In line with these properties, the interstellar medium (ISM) traced by the warm ionized gas is characterized by a high excitation and low gas-phase metallicity, with mean oxygen abundance about 0.2 times the solar value \citep[][]{Amorin+10,Amorin+12a,2012PASP..124...21H}, and a complex \ha\ kinematics with relatively high velocity dispersion \citep{Amorin+12b}. 

The extreme ionization properties shown by GPs, such as their large [\oiii]\,$\lambda\lambda$\,5007,4959\,\AA/[\oii]\,$\lambda$\,3727\,\AA\ (O32) ratios \citep{Jaskot+13}, have lead to suggest that the current starburst in GP-like galaxies proceeds under nearly "density-bounded" conditions, which may favor the escape of ionizing radiation, i.e. \lya\ and Lyman continuum (LyC) photons \citep{Jaskot+13, 2014MNRAS.442..900N, 2015A&A...576A..83S}. This was confirmed in the last few years in a series of studies based on observations performed using the Cosmic Origins Spectrograph (COS) on board the Hubble Space Telescope (HST) of the GPs rest-frame UV spectra: 100\% of the 11 GPs (and similar compact starbursts) at $z>0.3$ observed at rest wavelengths below the Lyman limit ($<912$\AA) shown direct indication of LyC leakage, with LyC escape fractions, $f_{\rm esc}$(LyC), spanning in a wide range of $\sim$2-70\,\% \citep{2016MNRAS.461.3683I,2016Natur.529..178I,2018MNRAS.474.4514I,2018MNRAS.478.4851I}. It is worth to be noted that these systems were selected by their high [\oiii]/[\oii] ratio, therefore a high fraction of LyC leakers was expected. Other compact starbursts at lower redshift where LyC detection is not directly accessible to observations, show indirect indication of leakage \citep{Alexandroff+15}. On the other hand, most GPs appear as \lya\ emitting galaxies, with a large fraction of them showing high \lya\ luminosities and EWs well in the range of high redshift galaxies  \cite[see e.g.][for a comparison]{2017ApJ...844..171Y}. They predominately show complex, narrow double-peaked, \lya\ line profiles and faint low-ionization absorption lines, suggesting low H{\sc i} column densities \citep{Jaskot+14,2015ApJ...809...19H,2015A&A...578A...7V,2017ApJ...844..171Y,Chisholm+17} and strong outflows \cite{Chisholm+15,Alexandroff+15,Heckman+16}. 

The above results are relevant in the context of cosmic reionization since recent observations suggest that the integrated properties of GPs are increasingly common at higher redshifts \cite[see e.g.][for a review about low mass galaxies at high redshift, and \citealt{Amorin+17} for a comparison with the GPs properties]{2014ApJ...784...58S,2016ARA&A..54..761S} and the first direct detections of LyC leakers in star forming galaxies at $z>3$ \cite[e.g.][]{2016A&A...585A..51D,2016ApJ...825...41V,2018MNRAS.476L..15V,2016ApJ...826L..24S} suggest that  young vigorous starburst episodes in low-mass galaxies may actually be a dominant source for cosmic reionization at $z>6$. 
However, the nature of the ionizing sources in Ly$\alpha$ emitters at these high redshifts is still under debate \citep{Dors+18}. There are several different interpretation of the scarce observational data of this kind of objects:
(i) Star Forming Regions with different contributions to the stellar populations that might include population III stars \citep[see e.g.][]{Sobral+15,Matthee+17} or not \citep[see e.g.][]{Bowler+17,Sobral+17}; (ii) Active Galactic Nuclei \citep[AGNs; see e.g.][]{Bowler+17,Laporte+17}; and (iii) Direct Collapse Black Holes \citep[DCBHs;][]{Agarwal+17}.
On the other hand, it is still not clear what are the necessary conditions driving the escape of ionizing photons. \citet{2018MNRAS.478.4851I}, suggest that a high ionization parameter is necessary but not sufficient condition for a large escape fraction based on the weak correlation found between O32 ratio and $f_{\rm esc}$(LyC). Instead, they found a strong anti-correlation with the peak separation of the blue and red \lya\ components, as predicted by radiation transfer simulations \citep{2015A&A...578A...7V}. It is therefore possible that a combination of properties, including age of the starburst, compactness (parametrized by high SFR surface densities), H{\sc i} covering fraction, H{\sc ii} region geometry and how this is shaped by gas kinematics, might favor the escape of a large fraction of ionizing photons.   
Whether the GPs show clear indication of strong outflows or not might be of paramount importance in the context of feedback process and the escape of ionizing photons.

Several studies have dealt with internal kinematics of violent starburst galaxies. \cite{2010Natur.467..684G} found that two thirds (11 out of 17) star forming galaxies in their sample exhibit kinematic signatures of rotating disks. They also conclude that feedback from the star formation process itself is the main driver for turbulence either from  stellar winds or supernovae \cite[see also][]{Green+14,Moiseev+12}. More recently the KMOS3D survey by \cite{Wisnioski+15,Wisnioski+18} explored the \ha\ kinematics of a large number of massive ($M_* > 4 \times 10^9 M_{\odot}$) starburst galaxies at z$\sim$1 and z$\sim$2. A vast majority of these systems are rotationally supported (i.e. rotational velocity larger than velocity dispersion). \cite{Alexandroff+15} presented observational evidence supporting a correlation between Lyman photons leakage and the compactness of the star-forming region itself.
The role of mechanical feedback driven by massive stars in compact starbursts is one of the above properties still not fully understood in relation with the ISM shaping and escape of ionizing photons. 

Using deep high dispersion spectroscopy of six GPs obtained using ISIS at the William Herschel Telescope (WHT), \citet{Amorin+12b} showed that their emission line profiles are not single Gaussians but a complex 
ensemble of independent kinematic components superposed on kpc scales, similarly to some other compact starbursts in the local universe \cite[see e.g.][]{Hagele+07,Hagele+09,Hagele+10,Hagele+13,Firpo+10,Firpo+11,2011ApJ...735...52B,2014MNRAS.442.3565C,Terlevich+14}. Amor\'in and collaborators showed that GPs present strong broad emission-line components with expansion velocities of several hundredths \kms\ or even higher,  which may account for up to $\sim$40\% of the total \ha\ luminosity. They also reported that these broad components are not only present in \ha\ but also in all the observed forbidden lines and in some cases show high relative velocities to the systemic one (up to 500 \kms), thus producing strong asymmetries in the line profiles. On top of the broad component, GPs generally show more than one narrower component, with relatively high intrinsic velocity dispersions of $\sim$\,30-80 \kms. 
While these narrow components appear to be kinematically resolved star-forming clumps associated to UV clumps only resolved by HST imaging, their high velocity dispersion points to a nearly turbulent ISM. 
In addition, the presence of strong broad emission is interpreted by \citet{Amorin+12b} as the imprint of energetic outflows from strong stellar winds and recent SNe. This scenario is consistent with the young star formation seen in most GPs and the presence of Wolf-Rayet features in some of them \citep{Amorin+12a}. 

Nevertheless, the lack of spatial resolution in long slit spectra precludes to investigate in more detail the spatial extent and distribution of such broad emission and its nature. For example, other processes such as the presence of faint AGN activity or turbulent mixing layers \citep[TML, e.g.][]{Binette+09}, that are difficult to rule out only using integrated spectra, could eventually be disentangled by means of Integral Field Spectroscopy (IFS). Moreover, IFS may help to study the association between the narrow kinematic components and spatially resolved star-forming clumps. Even if these high surface brightness regions are dispersion-dominated, using IFS we can analyze the GPs velocity field and try to find evidence of underlying rotation or signatures of merging. Previous attempts for similar analysis using IFS on a few GPs and similar objects have been limited by both spatial and spectral resolution and depth in the visible \citep{2017MNRAS.471.2311L}, or have been restricted to the NIR AO observations \citep{2010ApJ...724.1373G}.

\begin{figure}
\includegraphics[width=1\columnwidth]{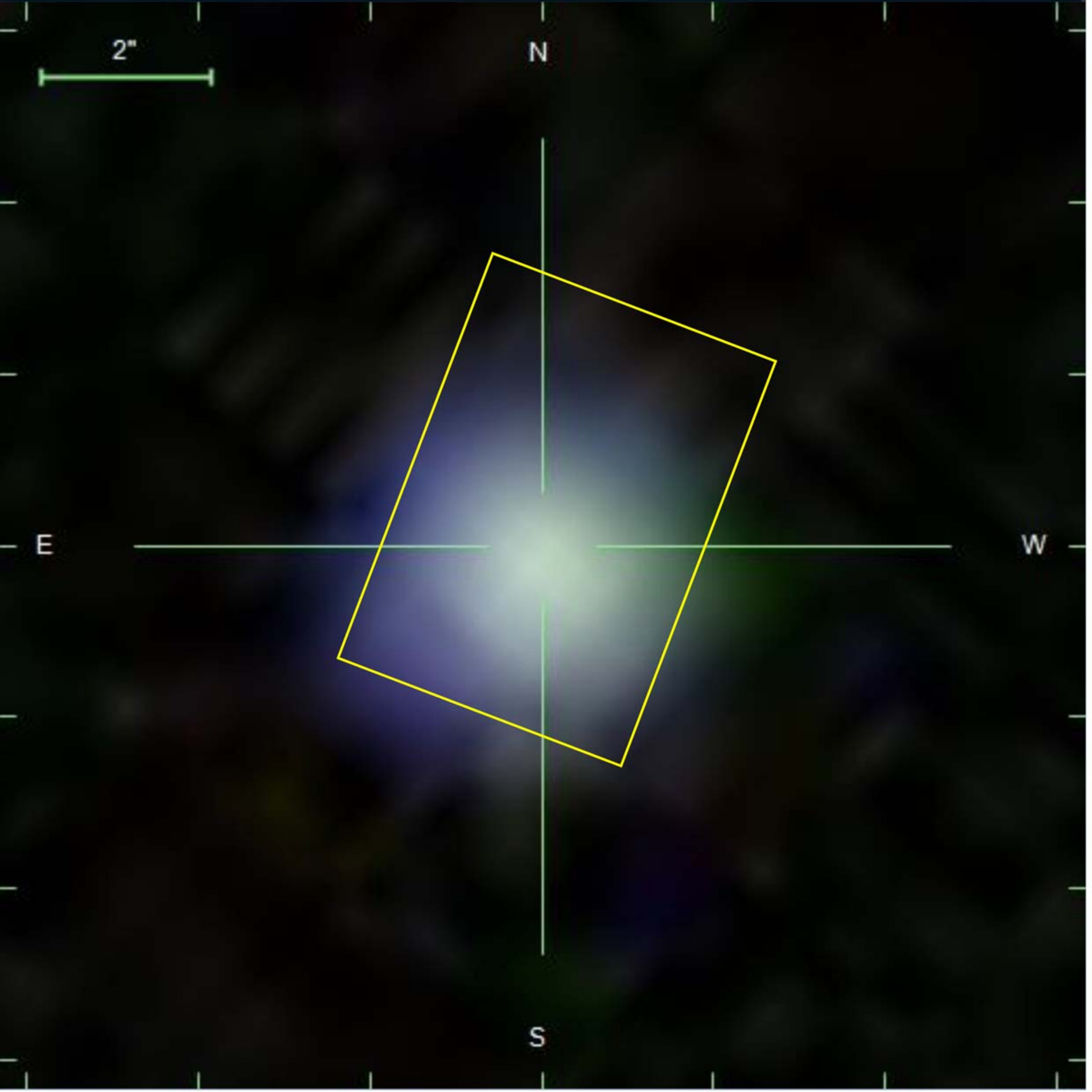}
\caption{Sloan Digital Sky Survey composite colour image of \ourobj with the observed GMOS-IFU field-of-view overimposed.}
\label{sloan}
\end{figure}

In this paper we present 3D spectroscopic data of SDSSJ083843.63+385350.5 (hereafter \ourobj), an emission-line galaxy located at $\alpha_{\rm J2000}$\,=\,08:38:43.6 and $\delta_{\rm J2000}$\,=\,38:53:50.0 (see Fig.\ \ref{sloan}) at a redshift of z=0.147. This galaxy is one of the brightest objects of a large sample of GPs selected by Amor\'in et al.\ (in prep.) from the SDSS-DR7 for detailed analyzes using low and high-dispersion spectroscopy. 
{\ourobj} shows a complex kinematic behavior in its SDSS spectrum. This galaxy was also included in samples of strong starbursts, \cite{2012MNRAS.421.1043S} confirmed its star forming nature from its location in a BPT diagnostic diagram \citep{Baldwin+81} and highlighted the presence of WR features in its spectrum. \cite{Shim+13} studied its properties as local universe analog to high redshift (z > 4) star forming galaxies. From these papers, basic spectral properties can be summarized: EW(H$\alpha$)\,=\,620\AA;  H$\alpha$ flux\,=\,2\,$\times$\,10$^{-14}$\,erg\,s$^{-1}$\,cm$^{-2}$, 
the \ha\ luminosity translates by the \citet{Kennicutt89} calibration to a star formation rate (SFR) of 22.5\,M$_{\odot}$\,yr$^{-1}$, assuming a continuous SF scenario since ~100 Myr, and logarithmic stellar mass $\log(M_*$/M$_\odot$)=9.573 (from a multi-wavelength photometry fitting).

Our aim is to disentangle spatially and kinematically the complex behavior previously observed in the emission lines of \ourobj.
In Section \ref{obs} we present the observations and data reduction, the data cube analysis is described in Section \ref{cube} and the results are shown in Section \ref{res}. The discussion and conclusions are presented in Section \ref{disc} and \ref{conc}, respectively.

\section{Observations and Reduction}
\label{obs}

Observations were carried out using the Gemini Multi-Object Spectrograph (GMOS) attached to the Gemini North Telescope in queue mode (Proposal Id:\,GN-2014B-Q-20; P.I.: G.\ H\"agele). \ourobj was observed in IFU 1-slit mode covering a field-of-view, (FOV) of 3.5\,$\times$\,5\,arcsec$^2$, sampled with 500 lenslets, plus 250 lenslets 60 arcsec offset for simultaneous sampling the background for sky subtraction. We used the R831 grating together with the OG515 filter to avoid second order contamination which yielded a spectral range from 6500 to 8500\,\AA\ (from 5550 to 7250\AA\ at the rest frame of the source) with a resolution R of 5100 for the H$\alpha$ line observed at 7500\,\AA. The wavelength resolution was derived from the emission-lines of the calibration lamps which present giving an average instrumental dispersion of $\sigma_{\rm inst}$\,$\sim$\,0.6 \AA, equivalent to 25\,\kms.
The grating angle was selected to ensure that the H$\alpha$ emission-line was observed.
Exposures were taken under excellent seeing conditions (average seeing of about 0.5\,arcsec), providing an excellent sampling with the 0.2\,arcsec IFU fiber size. The total observing time on source was 2.85\,h splitted in 6 equal exposures. Due to observing time constraints, only one baseline standard flux standard star (Hiltner\,600) was observed for flux calibration. 

Data was reduced using Revision 151 of the IFUDR GMOS package, based on the Gemini IRAF\footnote{IRAF: the Image
 Reduction and Analysis Facility is distributed by the National Optical
 Astronomy Observatories, which is operated by the Association of
 Universities for Research in Astronomy, Inc. (AURA) under cooperative
 agreement with the National Science Foundation (NSF).} 
 package, which was provided by James Turner and Bryan Miller via the Gemini Data Reduction User Forum. A custom script adapted to our dataset was used to produce a flux and wavelength calibrated data cube for the field covered in our pointing. 

\section{Data Cube Analysis}\label{sec:analysis}
\label{cube}

The Gaussian fitting of the emission line profiles of more than 400 spectra obtained in individual spaxels is an interesting challenge on its own. Almost every analysis software, such as \textit{ngauss} in IRAF or  \textit{PAN}\,\footnotemark\ in IDL, rely on manually providing a set of initial guesses for the key parameters that require substantial interaction with the user.
 This turns out crucial as the least square minimization relies strongly on the goodness of the initial guess when the number of components increase and with them, the number of local minima of the $\chi^2$ estimator. In our case, we are handling data for a whole galaxy, where complex line profiles within a galactic velocity field are expected \citep[see e.g.][]{Amorin+12b}. There is no single `initial velocity' value for all spaxels and the relative velocity of components are expected to vary throughout the GMOS-IFU field. A good approach is to obtain a first order approximation of the global velocity field present in the observed area and analyze the presence of multiple components that could arise from deviations from global kinematics on top of that average velocity field. Furthermore, the overall flux, another quantity requested as initial guess for the profile, changes dramatically from the bright center toward the galaxy outskirts.
\footnotetext{The package PAN (Peak ANalysis; Dimeo 2005, PAN User Guide ftp://ftp.ncnr.nist.gov/pub/staff/dimeo/pandoc.pdf) adapted and modified by \citet{Westmoquette+07} for astronomical requirements}

To achieve this we built the Python code MultiGauss3D (\mgd) that analyzes every individual spectrum of the data cube. The actual fitting relies on the Non-Linear Least-Square Minimization and Curve-Fitting (LMFIT) package \citep{Newville+14}. \mgd applies an iterative process: based on the SDSS global spectrum of the galaxy we estimated an average velocity and velocity dispersion for the galaxy, which was used as a starting point for a single Gaussian fit to each emission line profile. Although the amplitude varies significantly over the field, this has proven to be very robust and quickly converges from the initial guess to a first order approximation that traces the global velocity behavior and scales the intensity of the initial guess appropriately. This output is used to derive a second component, split from the first one, where only the fluxes of each components are scaled, with a 2/3 and 1/3 factor for the original and cloned versions, respectively, and using for both components the same kinematic result derived for the single Gaussian fit. The fitting procedure is repeated with this new set of initial guesses and the new output is analyzed to check if there is an actual improvement in the fit from the previous model. This check is done using the Akaike Information Criterion indicator \citep[AIC,][]{Akaike74} which allows us to compare competing models with different number of Gaussian components and, therefore, different number of free parameters. Following \cite{Wei+16} we consider $\Delta\mathrm{AIC} > 10$ as `very strong' evidence for a real need of an increase in the number of Gaussian components. For spaxels where the increase in number of components is justified by $\Delta\mathrm{AIC}$, the code adds a further component following the same criterion as described above and performs a new fit. This procedure is repeated until the \textit{(N+1})th component fails the Akaike test and the fit is settled to \textit{N} components. Visualization plots and graphs are produced within the code with the aid of the Python routines \texttt{display pixels} and \texttt{plot velfield} devised by Michele Cappellari and presented in \cite{2006MNRAS.366..787K}.

\begin{figure*}
\includegraphics[width=2\columnwidth]{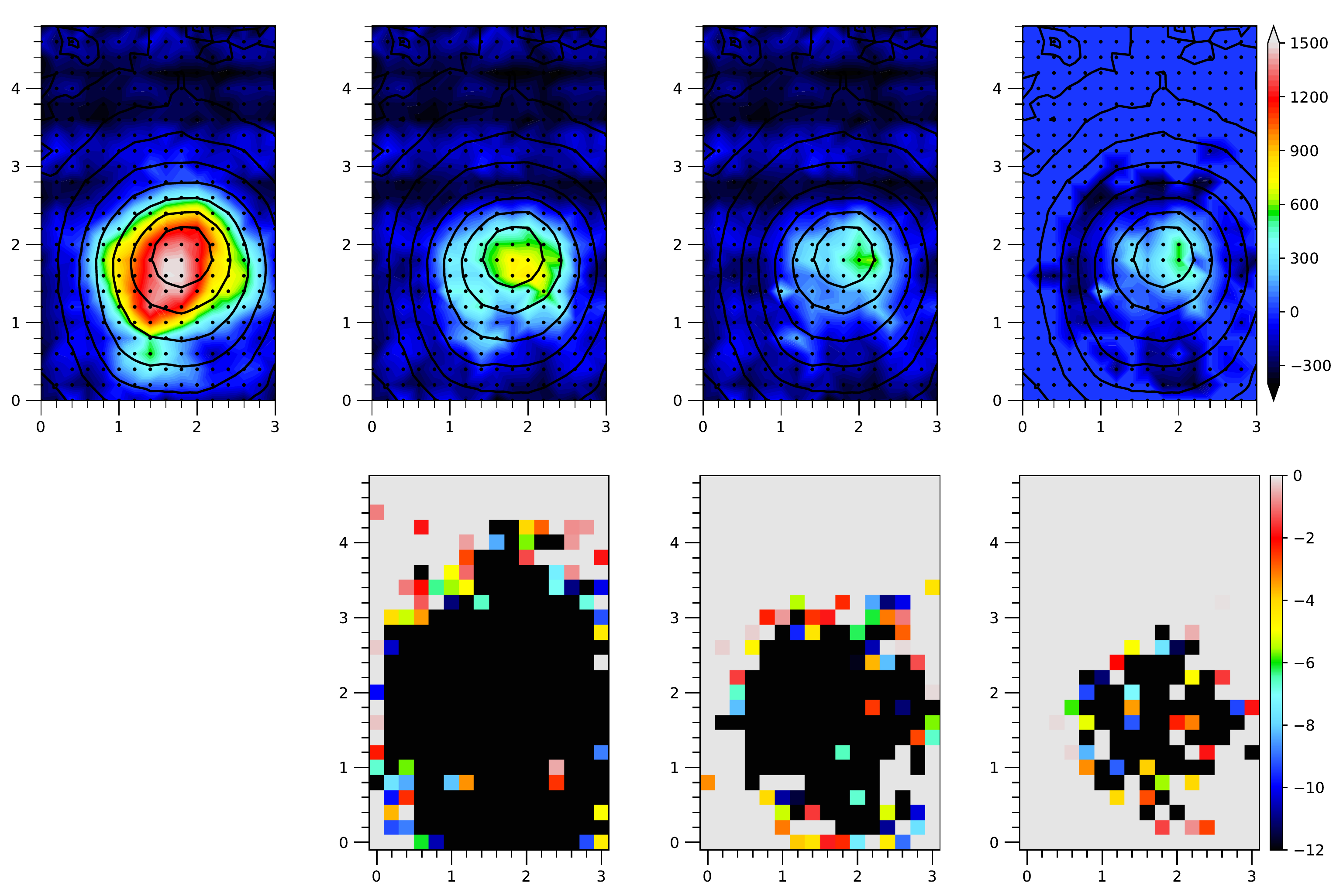}
\caption{Graphical breakdown of the fitting procedure. X and Y axes are arc-seconds from the bottom left corner of the fiber array. Top line: From left to right the evolution of the Akaike Information Criterion as the number N of Gaussian components is increased from 1 to 4. Bottom line: The difference between N and (N-1) AIC, where the black spaxels indicate where $\Delta\mathrm{AIC}$ supports the increase in number of components. Rightmost bottom panel reveals that the number of black spaxel area is smaller than 20\% of the total area.  See full description in the text.}
\label{Akaike}
\end{figure*}

The turnout of the Akaike test includes a combination of actual presence of a number of kinematical components and the signal to noise ratio on each spaxel. As \ourobj is a compact starburst, the flux received by each spaxel decreases rapidly from the galaxy center which in turn degrades the S/N ratio. This can be readily seen in Fig.~\ref{Akaike} when comparing the areas covered by the different successful models. A decision has to be made on the largest meaningful number of components to perform a kinematical analysis throughout the galaxy. A balance between kinematical detail and area needed to detect behavior pattern must be kept, so in our analysis we decided to 
discard the four component analysis as
the number of spaxels available to the analysis were less than 20\% of the initially surveyed area.

\begin{figure}
\centering
\includegraphics[width=0.7\columnwidth,height=0.7\columnwidth]{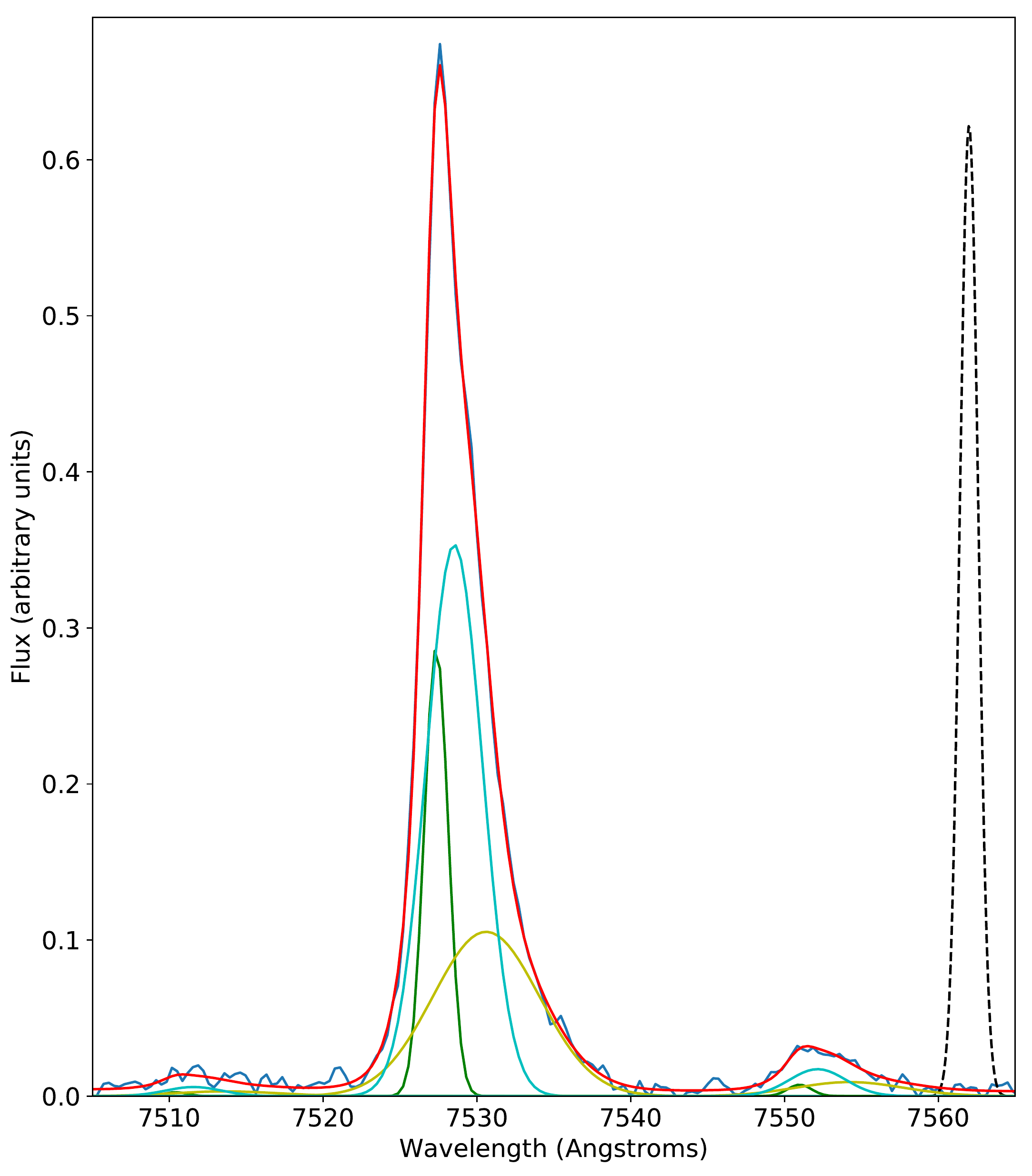}\\
\includegraphics[width=0.7\columnwidth,height=0.7\columnwidth]{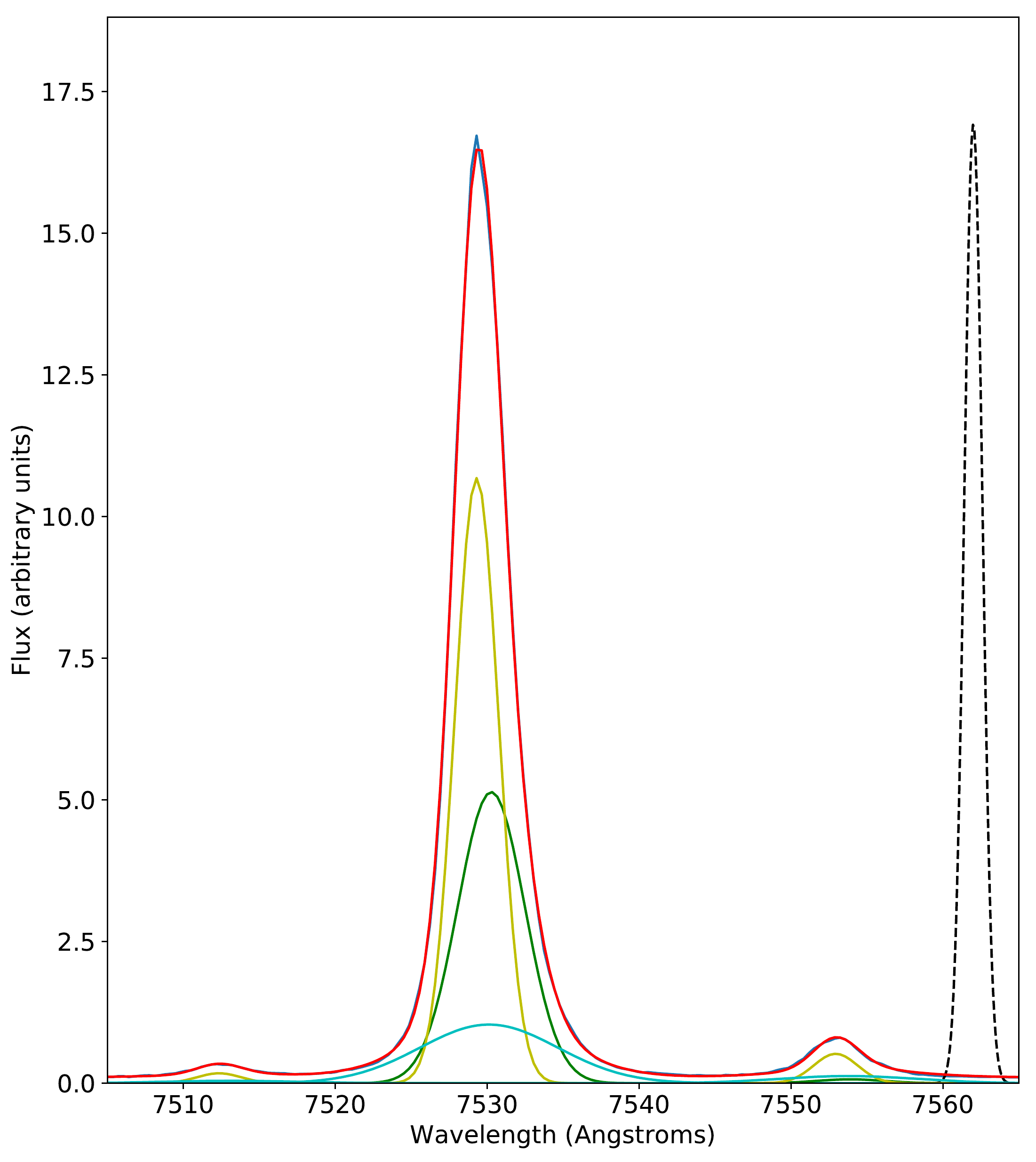}\\
\includegraphics[width=0.7\columnwidth,height=0.7\columnwidth]{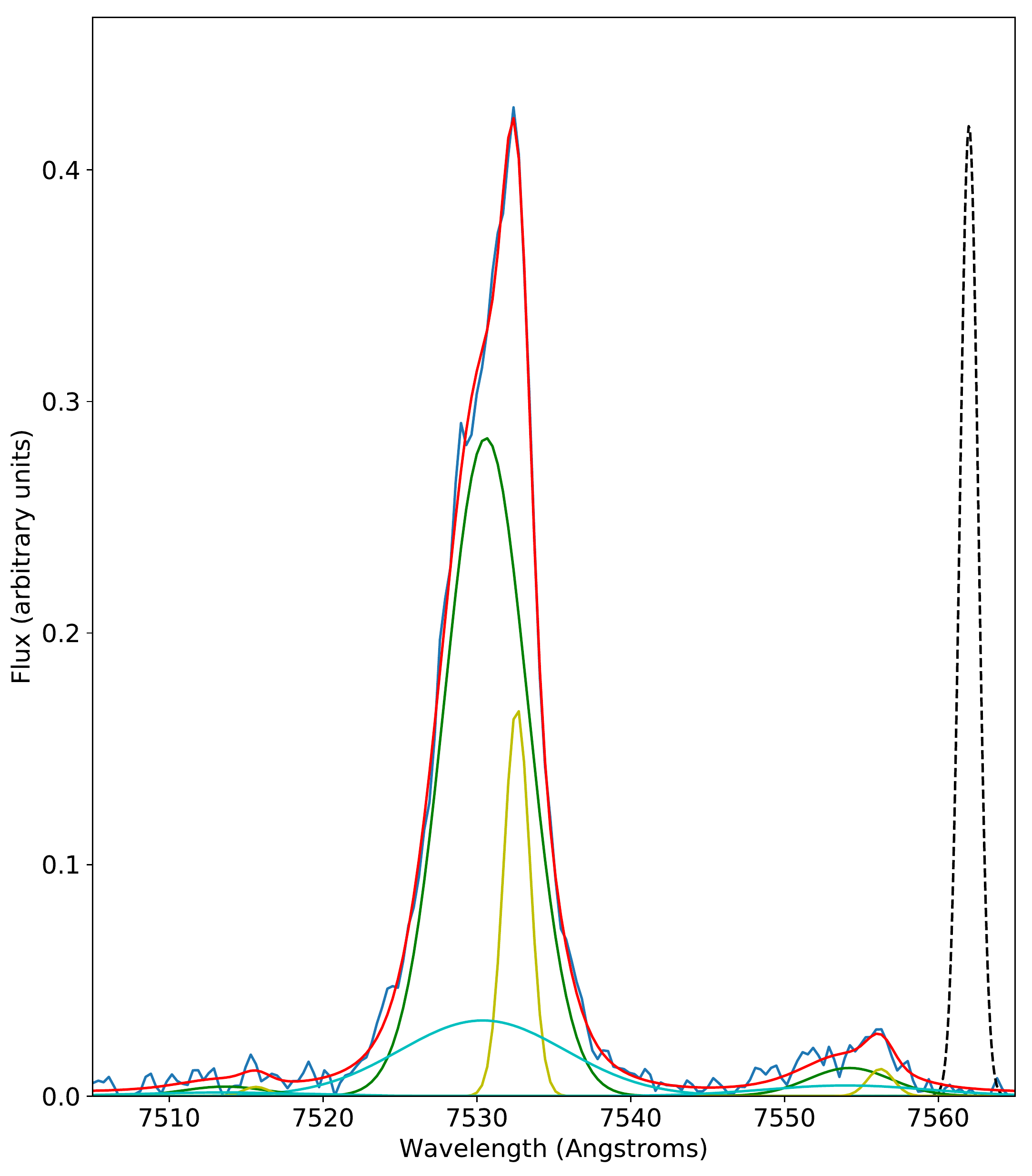}
\caption{Examples of the simultaneous fitting of three components to the H$\alpha$ and the [\ion{N}{ii}]\,6548,6584\,\AA\ emission lines, highlighting the strong variations detected for three different spaxels separated by about one arc second (five spaxels) in the y direction along the array. For each line the three different kinematical Gaussian components and their sum (in red color) are shown. For comparison purposes, a scaled instrumental profile ($\sigma=0.6$ \AA\ ) is shown in dashed black.}
\label{Examplefit}
\end{figure}

For our GMOS spectra on \ourobj, the largest meaningful number of Gaussian components was found to be three.
Among the emission-lines present in our spectra, the H$\alpha$ recombination line is by far the strongest one and was therefore used to derive the detailed map of the ionized gas kinematics. [\ion{N}{ii}] emission-lines at 6548, 6583 \AA\, lie on the wings of the broadest H$\alpha$ component and were therefore included in the fit to improve the determination of the adjacent continuum values (see Fig.\ref{Examplefit}). As these [\ion{N}{ii}] lines are intrinsically weak for this kind of objects, it is more reliable to restrict the number of free parameters involved in the fit. 
Since it is involved roughly the same energy to produce the H$\alpha$ and the [\ion{N}{ii}] emission-lines, they are expected to be approximately originated in the same spatial zone, hence, 
the Gaussian center and dispersion were fixed to the H$\alpha$ ones and only the flux was fit through a scaling factor \citep[see details in][]{Firpo+10,Firpo+11,Hagele+12}, bearing also in mind that there is a theoretical relation among the fluxes of the [\ion{N}{ii}] lines \citep[$I$(6584)\,=\,2.95\,$\times$\,$I$(6548),][]{Osterbrock+06}. The total number of free parameters on each fit spanned from six (3 H$\alpha$ Gaussian parameters, 2 continuum parameters and 1 scaling factors for the [\ion{N}{ii}] lines) when fitting a single Gaussian, to eighteen (12 for H$\alpha$, 2 for continuum and 4 for the [\ion{N}{ii}] lines) when fitting up to four kinematical components. The LMFIT routine allows to set lower and/or upper limits for the parameters to be fit, which we used to set a minimum $\sigma$ value for each component larger than the measured instrumental width and force the task to fit emission-lines by setting the minimum Gaussian flux to be larger than zero. These limits allow to constrain the minimization procedure to deliver meaningful results, the only drawback is that when a parameter reached its boundary value, there was no reliable estimation of this parameter (the errors are set to zero, and this estimation is not taken into account for subsequent maps and estimations). 

\begin{figure*}
\includegraphics[width=0.9\textwidth]{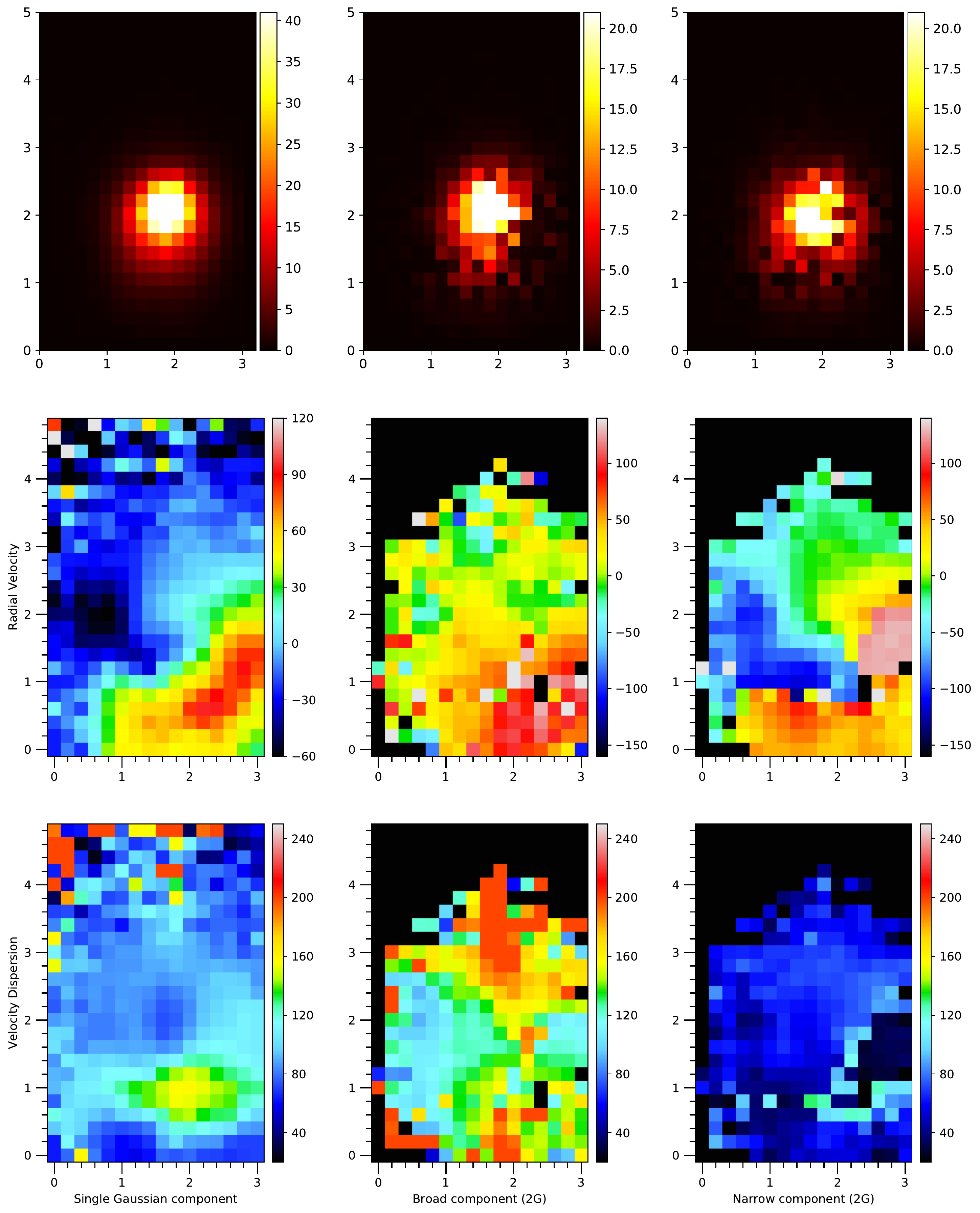}
\caption{H$\alpha$ flux and kinematics maps for Single Gaussian (left column) and two Gaussian component (middle and right columns for broad and narrow components respectively). X and Y axes are arc-seconds from the bottom left corner of the fiber array. Top row shows flux distribution in units of $10^{-15}$ erg s$^{-1}$ cm$^{-2}$. Second row shows the corresponding radial velocity maps and third row displays the velocity dispersion maps. All velocities and velocity dispersions are shown in \kms.
}
\label{Single_2G_maps}
\end{figure*}

\begin{figure*}
\includegraphics[width=0.8\textwidth]{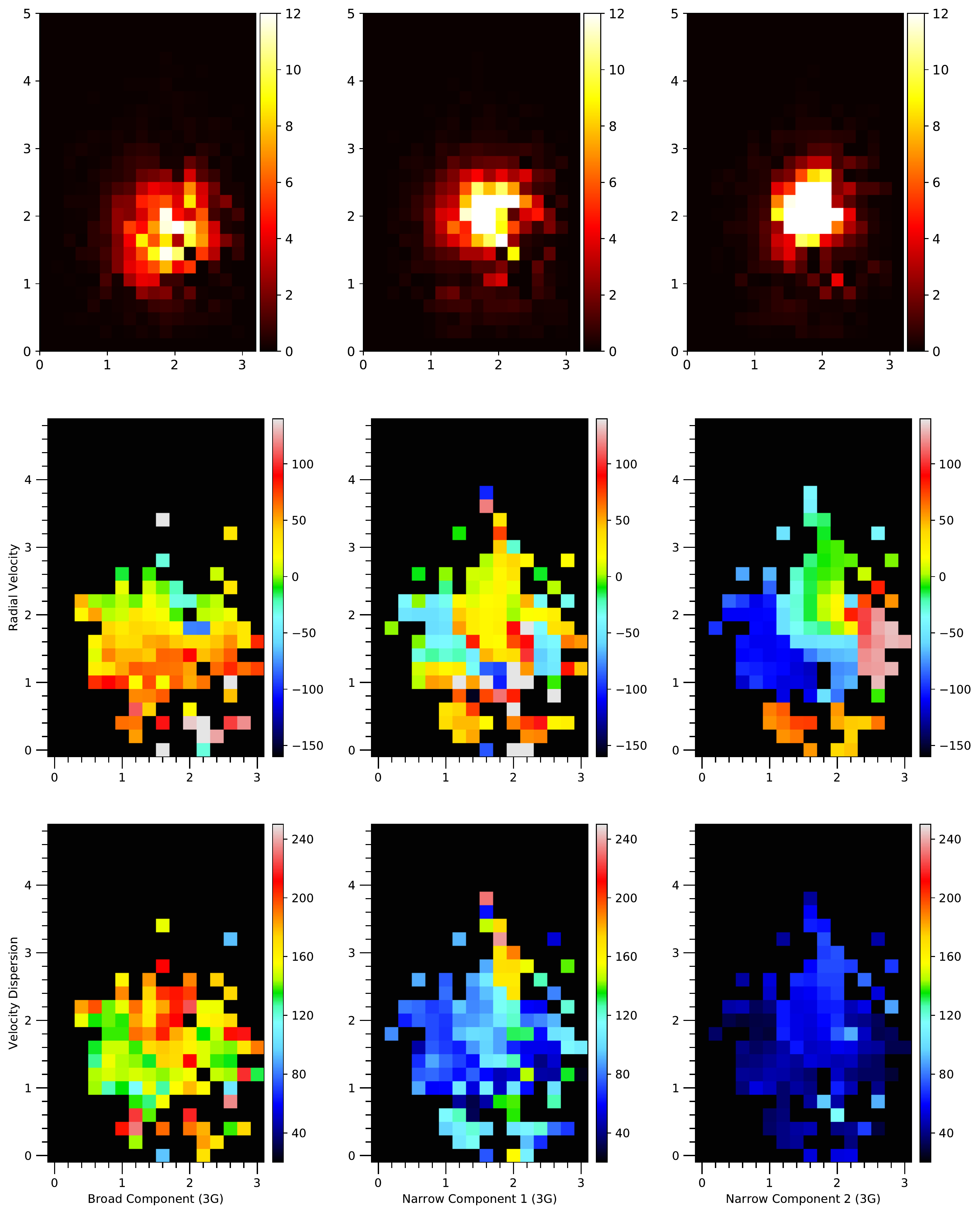}
\caption{Same as figure \ref{Single_2G_maps} for the three Gaussian component fits} 
\label{3G_maps}
\end{figure*}

Besides the $\Delta$AIC that accounts for the statistical significance, we also checked for other flags that highlight potential problems from the fit output. These include large uncertainties in the parameters determination or when one of the parameters reaches the limit set in the constraints, such as when the narrow component width matches the instrumental profile width.  
This was achieved by building masks that flagged individual spaxels which met any specific problem
which were later combined accordingly to build the overall mask to apply when analysing spatial behavior of any given parameter.


\section{Results}
\label{res}

The possibility of fitting multiple components on the H$\alpha$ emission-line profiles of the data cube allows us to perform a much more detailed analysis of the gas kinematics, as the behavior differs dramatically from what can be inferred from fitting a single Gaussian profile or an integrated spectrum.
This can be readily seen in the radial velocity maps shown in the second row of Fig.~\ref{Single_2G_maps}, which shows the different maps derived from a single Gaussian fit (left panel) and what can be obtained from a more comprehensive analysis using two Gaussian components (middle and right panels). Radial velocity maps for three components analysis are included in the second row of Fig.~\ref{3G_maps}. The corresponding intrinsic velocity dispersion ($\sigma$; where $\sigma^2$=$\sigma_{\rm g}^2-\sigma_{\rm i}^2-\sigma_{\rm t}^2$ being $\sigma_g$ the velocity dispersion from the fitted Gaussian and $\sigma_i$ and $\sigma_t$ the instrumental and the thermal broadening ones, respectively) maps are shown in the third rows of Figs.\ \ref{Single_2G_maps} and \ref{3G_maps}. For the two and three Gaussian fits, the components are displayed in decreasing velocity dispersion order. For the last case these components are hereafter labeled BC, NC1 and NC2, respectively.

Inspection of the velocity panel for the single Gaussian fit suggests evidence of galaxy rotation, which would be difficult to reconcile with the velocity dispersion which peaks off galaxy center 
(left column, bottom map in Fig.~\ref{Single_2G_maps}). The inclusion of a second Gaussian component already reveals the presence of two distinct kinematical components: a narrow component that suggests the presence of a rotating feature (Fig.~\ref{Single_2G_maps} right column, middle map) and a broad component with no traces of rotation at all (Fig.~\ref{Single_2G_maps} middle column, middle map). On the other hand, while the velocity dispersion of the narrow component shows a flat distribution (Fig.~\ref{Single_2G_maps} right column, bottom map), the broad component exhibits an increase in its velocity dispersion away from the center along the y-axis (Fig.~\ref{Single_2G_maps} middle column, bottom map). This already shows the strong impact that fitting more than one component has on the interpretation of the ionised gas kinematics.
After being able to split the observed profile in one broad and two relatively narrower components, a different scenario arises. The broad component, BC, has a complex behavior: on top of a smooth velocity distribution, the spaxels that exhibit the largest velocity dispersion (Fig.\ \ref{3G_maps} left column, bottom map) also seem to show the largest radial velocity deviations from a systemic one (either blueshifted or redshifted) (Fig.\ \ref{3G_maps} left column, middle map).
These spaxels are also located opposite to the galaxy center which suggests the presence of an outflow (nearly bi-conic in projection) with high $\sigma$ approximately in the `y-axis' direction. On the other hand, the lower velocity dispersion components, NC1 and NC2, show different behaviors. NC2, the narrowest component, shows another arranged-motion behavior but with much larger amplitude than the one exhibited by BC, and extended along the `x-axis' (Fig.\ \ref{3G_maps} right column, middle map). It seems to trace a disk rotation with an almost flat velocity dispersion with a small increment of $\sigma$ towards the centre. It is noticeable that the velocity field of the other narrow component, NC1, (Fig.\ \ref{3G_maps} middle column, middle map) resembles the velocity pattern of NC2 with smaller deviations and its velocity dispersion (Fig.\ \ref{3G_maps} middle column, bottom map) resembles the general behavior of BC. 

A possible scenario to explain this behavior is that NC1 is originated in the external layers of the gaseous material that seems to follow an overall rotational movement (NC2) and, therefore, both components show similar velocity patterns. The similarities between NC1 and BC velocity dispersion maps
could be the result of the interaction of the outflow that originates BC with the rotating gas. This scenario is similar to the turbulent mixing layers scenario proposed by \cite{Westmoquette+09} to interpret the optical structure observed in the inner zone of M\,82.

The combination of these distinct kinematical components produces the effect observed when analyzing a single Gaussian profile. 
Focusing on the velocity dispersion maps, the off center maximum has disappeared and the profile broadening extends from the center towards the upper and lower edges. This pattern is similar and more evident in BC and NC1. 


At a redshift estimated distance of about 600 Mpc, the scale is roughly 2.5\,kpc per arcsec so our high velocity dispersion regions have sizes in the kiloparsec range. 
The single Gaussian fit shows an average velocity dispersion in the range of 80 - 90 km\,s$^{-1}$ (peaking up to $\sim$160\,km\,s$^{-1}$). Comparatively, BC shows higher velocity dispersion values with an average between 140 and 150\,km\,s$^{-1}$ peaking up to 240\,km\,s$^{-1}$.
Such a behavior has been detected in nearby starbursts, such as He\,2-10 \citep{Cresci+17}. NC2, in turn, presents more extreme radial velocity amplitudes, with about 250 km\,s$^{-1}$ shifts projected in radial velocity. 

\section{Discussion}
\label{disc}

The performance shown by the multiple component fit procedure confirms its validity as a tool for disentangling complex kinematical behavior with seeing-limited IFS observations. Analysing the results obtained from a single Gaussian component fit, \ourobj does not satisfy almost any of the conditions set by \cite{Wisnioski+15} to be classified as a rotating system. The possibility of splitting the observed profiles in different components reveals the presence of a relatively low dispersion gas (NC2) which can be associated with a rotating disk. The broader components (NC2 and BC) show no structured motions, although the orientation of the outflow cone structure, perpendicular to the alleged disk would match what is observed in starburst galaxies such as M82. \cite{1995A&A...293..703M} with longslit data on M82 had already shown the kinematical signature of strong outflows triggered by supernovae explosions in a direction perpendicular to the rotating disk and \cite{2013MNRAS.428.1743W} found within their GMOS mosaic a trend in which their broad component increased its width from the disk outwards, although their spatial coverage does not reach the galaxy outer region. Our GMOS-IFU data for \ourobj shows a similar behavior, with $\sigma$ of the BC increasing outwards within the conic region perpendicular to the rotating disk with a projected size of about 7\,kpc at the distance of this object. On the other hand, NC1 could be originated in the external layers of the gaseous material that seems to follow the overall rotational movement (NC2) and, therefore, both components show similar velocity patterns. Moreover, the velocity dispersion map of NC1 is similar to the one of BC, that could be the result of the interaction of the outflow that originates the broad component with the rotating gas.
Thus, NC1 might be associated with the mixing layer component such as those described by the scenarios proposed by \cite{Westmoquette+09} and \cite{Binette+09} for M82 and NGC\,2363, respectively.
The fact that the peak of the H$\alpha$ emission-line fluxes of NC1 and NC2 are almost spatially coincident and the BC peaks about 0.45\,arcsec to the south (see Fig.\ \ref{sloan} and upper panels of Fig.\ \ref{3G_maps}) also supports this idea.

In the local universe (z\,<\,0.1), the Blue Compact Dwarf galaxy ESO\,338-04 shows two outflows extending by about 3\,kpc each (projected distance), towards the North and the South, that could be originated inside or at the edge of the bubble created by the stellar winds and supernova feedback from a super star cluster as suggested by \citet{Bik+18}. These authors argued that the mechanical energy that created the bubble also heated and thermalized the gas inside it, and Rayleigh-Taylor instabilities made it to burst, originating the outflows seen at the galactic scale, in consonance with the standard model of galactic winds \citep[see e.g.][]{Chevalier+85,Heckman+17}.
A similar scenario could explain the bi-conical 7\,kpc (projected distance) outflow seen in \ourobj.

Although the combination of rotating disk and outflow kinematics has already been detected in other violent starburst galaxies, the rotating disk is far from dominating the integrated profile. How can we match this with what has been observed, for instance in the KMOS3D survey? A possible explanation can be found when comparing the SFR properties of \ourobj with other starburst galaxies (see discussion in Sect.\ \ref{intro}). \cite{Shim+13} estimated a SFR of 22\,M$_{\odot}$ yr$^{-1}$ using the H$\alpha$ flux of 4.56$\,\times\,$10$^{-14}$\,erg\,s$^{-1}$\,cm$^{-2}$ provided by the SDSS data base and applying an ad-hoc aperture correction factor of 2. However, using the same expression given by \citet{Kennicutt89}, we have estimated a SFR of about 18\,M$_{\odot}$ yr$^{-1}$ directly measuring the H$\alpha$ flux (=\,6.14$\,\times\,$10$^{-14}$\,erg\,s$^{-1}$\,cm$^{-2}$) from the same SDSS spectrum and applying a reddening correction c(H$\beta$)\,=\,0.26 calculated using the Balmer decrement taking into account the H$\beta$ emission-line and a theoretical value of 2.86, assuming the Galactic extinction law of \cite{Miller+72} with R$_v$\,=\,3.2.
We find that \ourobj is in the WISE point source catalog and we can derive its SFR from its observed mid-IR magnitudes. Following \cite{Brown+17} we will focus our comparison using the luminosity estimated in the W3 bandpass centered in 22 $\mu$m (L22) which displays the best correlation with other independent estimators. Several regressions for different SFR estimators have been derived recently by \cite{2013ApJ...774...62L}, \cite{Wen+14}, \cite{2016MNRAS.461..458D} and \cite{CatalanTorrecilla+15} for different samples of galaxies. While our estimation of the SFR for \ourobj from its
H$\alpha$ luminosity is of 18 M$_{\odot}$ yr$^{-1}$, all determinations from L22 (including those combining it with the H$\alpha$ luminosity) yield values in the range of 29-66 M$_{\odot}$ yr$^{-1}$ pointing towards a large percentage of UV missing photons.
We also identified \ourobj in a GALEX catalog of UV sources by \cite{2011Ap&SS.335..161B} and derived a SFR of about 1\, M$_{\odot}$ yr$^{-1}$ from the FUV and NUV magnitudes. Although quite lower than those derived from H$\alpha$ and MIR, it is somewhat expected that dust attenuation plays an important role \citep{Lee+09}. This high SFR combines with its low stellar mass \cite[$3.74 \times 10^9$\,M$_{\odot}$,][]{Shim+13} yielding a specific star formation rate (sSFR) estimated from the H$\alpha$ flux of about 5 Gyr$^{-1}$, being among the largest values estimated for the KMOS3D sample \citep{Wisnioski+15}. Moreover, the sSFR taking into account the L22 WISE data is in the range of 6-18 Gyr$^{-1}$, much larger than the strongest SFR galaxy in KMOS3D. Furthermore, \cite{Wisnioski+15} presented a correlation between the ratio of turbulence and rotation kinematical signatures ($v_{\rm rot}/\sigma_0$) and sSFR. Our finding that turbulence dominates the overall kinematics can be explained as a $\sigma_0 > v_{\rm rot}$ condition, provided \cite{Wisnioski+15} correlation holds for sSFR values as large as the one found for \ourobj.

An estimation of the rotational velocity can be made assuming that the amplitude of the velocity of the narrower component along the disk (NC2) is due to rotation, providing a $v_{\rm rot}$ lower estimate at around 100 \kms\ (see Fig.\,\ref{3G_maps}, right column middle map). Assuming the \citet{TullyFisher77} relation and its calibration provided by \cite{McGaugh12}, we estimated a mass of about $\log(M_{\rm rot}$/$M_\odot$)=9.67. Taking into account the errors in the calibration and in our velocity estimations, this value is consistent with the photometric estimation [$\log(M_*$/$M_\odot$)=9.573] performed by \citet{Shim+13}. On the other hand, velocity dispersions for BC lie in the order of the 200 \kms noticeably larger than the former, although the observed velocity for the narrow component might differ from the actual rotational velocity due to projection effects. It is hard to assess the inclination angle of the disk just from its kinematical features. Within the outflow scenario it can be seen in the work by \cite{2016MNRAS.460.2731C} that maximum outflow velocities lie in the range of a few hundred \kms. The maximum radial velocity estimated for BC is almost 100 \kms\ at the outer regions, well in the range proposed in this scenario. The broad emission component shows a full width at zero intensity (FWZI; which characterises the terminal velocity of the outflow) in the range of 900 - 1400 \kms, in agreement with the values found by \citet{Amorin+12b}.

The complex gas kinematics shown by \ourobj is both qualitatively and quantitatively similar to what was found by \citet{Amorin+12b} for a small sample of GPs using high resolution spectroscopy. Hence, we can 
link our results to the hypothesis suggested by these authors that presented GPs as rapidly assembling galaxies. Regarding the broad components, from the MOSFIRE Deep Evolution Field (MOSDEF) survey of 211 star-forming galaxies at z$\sim$1.37-3.8, \cite{Freeman+17} found a significant (at $>3\sigma$) broad component in about 10\% of the sample, even though this component is present in all their stacked spectra. They also found a correlation between the detection of the broad component and the signal-to-noise ratio of their spectra, which, as they remarked, implies a dependence in their ability to detect the broad components on the quality of the data. Thus, the kinematical properties derived for \ourobj and the GPs previously studied by \citet{Amorin+12b} seem to be somehow similar to those showed by clumpy star-forming galaxies at z$\sim$1-4.

The large differences in SFRs derived from UV and optical and those estimated from IR fluxes pointed out in the analysis described above also suggests that a large number of ionizing LyC photons are not being detected through recombination lines and are therefore escaping from the galaxy. This is in agreement with recent studies that detect LyC leakage in all GPs 
\citep{2016MNRAS.461.3683I,2016Natur.529..178I,2018MNRAS.474.4514I}.
Our kinematical analysis shows large velocity dispersions detected in the broad component and therefore it provides straightforward evidence that supports the scenario where dense interstellar material is being shredded by powerful outflows driven by the strong massive star winds and/or energetic outflows from supernovae events. 
These high velocity dispersions also supports the idea of the escape of Ly photons, in consonance with the results found by \citet{Herenz+16,Herenz+17} that claimed that there is a link between turbulence in the interstellar medium in SFRs and the escape of Ly radiation (Ly$\alpha$ and continuum).

In a recent work, \cite{Orlitova+18} performed an exhaustive analysis of the Ly$\alpha$ emission-line profiles for a sample of twelve  GPs observed using HST-COS and applying numerical radiative transfer models \cite[see][]{Verhamme+06}. Orlitov\'a and collaborators assumed an expanding and homogeneous spherical shell for their model geometry, composed by uniformly mixed neutral hydrogen and dust. In their very thorough study, they combined these model results with ancillary optical and ultraviolet data with the aim to constrain the model parameters. Regarding the Ly$\alpha$ profile model, they assumed that Ly$\alpha$ and H$\beta$ emission-lines are formed by the same recombination mechanism and hence the intrinsic Full With Half Maximum (FWHM) has to be the same for both of them. On this basis they fitted the H$\beta$ emission-line profiles using the low/intermediate spectral resolution (R$\sim$2000) SDSS spectra of their GP sample assuming a two components Gaussian function with a dominant narrow component and broad one. Due to the low resolution of their data these authors assumed a 100\,km s$^{-1}$ uncertainty for their measurements of the FWHM of H$\beta$, which is slightly high even for this kind of fit even using low/intermediate spectral resolution data. These authors had to use an input model-line profile broader than the one estimated from the SDSS data to be able to reproduce the observed Ly$\alpha$ profile. They also pointed out that high resolution spectroscopy could potentially help to better understand other inconsistencies between model results and the observational data.


As was highlighted by \cite{Orlitova+18}, a deeper analysis of the implications of the very complex and highly variable spatial profiles of the Balmer emission-lines in GPs is necessary for a better interpretation of the Ly$\alpha$ emission-line profiles and to infer the physical conditions of the medium that are affecting them. To improve the accuracy of this kind of analysis it is also necessary to estimate the effects of the reddening for which we must obtain observations of other Balmer lines, mainly H$\beta$, in the blue range of the optical spectrum. The SDSS spectra are not enough for this purpose since it would be needed an IFU-echelle combination that, in turn, needs a great collecting area as the one would be provide for the next generation of giant telescopes as e.g.\ the Giant Magellanic Telescope (GMT).

\section{Conclusions}
\label{conc}
The characterization of ionized gas kinematics within massive violent starbursts requires a spectroscopic analysis that matches its complexity. In this work we have presented a thorough analysis of Integral Field Spectroscopy data on the strong star forming galaxy SDSSJ083843.63+385350.5. 
We have
shown that single Gaussian profile fitting techniques might miss to detect the underlying dynamics within the galaxy. Our multiple component fitting unveiled the presence of: (i) a rotating signature probably linked to a disk feature, (ii) a violent outflow which extends to several kiloparsecs perpendicular to the disk plane, and (iii) an intermediate component that can be identified with a turbulent mixing layer already identified in closer starburst galaxies. We discuss the link that the outflow signature might have with the specific star formation rate and propose that this outflow can be responsible for shredding the interstellar material and allowing the escape of Lyman photons detected in these type of galaxies.

\section*{Acknowledgments}

We acknowledge our anonymous referee for a thorough revision of the manuscript and the useful comments and suggestions that helped us improve the final version of this paper.
VF acknowledges support from CONICYT Astronomy Program-2015 Research Fellow GEMINI-CONICYT (32RF0002). OLD is grateful to FAPESP and CNPq. ACK thanks the CNPq support. 
This work is based on observations obtained at the Gemini Observatory, which is operated by the Association of Universities for Research in Astronomy, Inc., under a cooperative agreement with the NSF on behalf of the Gemini partnership: the National Science Foundation (United States), the National Research Council (Canada), CONICYT (Chile), Ministerio de Ciencia, Tecnolog\'{i}a e Innovaci\'{o}n Productiva (Argentina), and Minist\'{e}rio da Ci\^{e}ncia, Tecnologia e Inova\c{c}\~{a}o (Brazil).

\bibliographystyle{mnras}
\bibliography{biblio.bib}

\clearpage

\clearpage

\clearpage

\bsp	
\label{lastpage}
\end{document}